\documentclass[10pt,conference]{IEEEtran} 
\usepackage{times}

\usepackage[colorinlistoftodos]{todonotes} 

\usepackage{wine} 

\begin{document}
\pagenumbering{gobble}

\title{\textbf{\Large semantify.it, a Platform for Creation, Publication\\[-1.5ex]and Distribution of Semantic Annotations}\\[0.2ex]}

\author{
\IEEEauthorblockN{~\\[-0.4ex]\large Elias K\"arle, Umutcan \c{S}im\c{s}ek and Dieter Fensel\\[0.3ex]\normalsize}
\IEEEauthorblockA{STI Innsbruck, University of Innsbruck,\\
Technikerstrasse 21a, 6020 Innsbruck, Austria\\
\{elias.kaerle, umutcan.simsek, dieter.fensel\}@sti2.at}}

\newcommand{\ts}{\textsuperscript}

\maketitle

\begin{abstract}
The application of semantic technologies to content on the web is, in many regards, important and urgent. Search engines, chatbots, intelligent personal assistants and other technologies increasingly rely on content published as semantic structured data. Yet, the process of creating this kind of data is still complicated and widely unknown. The \textit{semantify.it} platform implements an approach to solve three of the most challenging question regarding the publication of structured semantic data, namely: a) what vocabulary to use, b) how to create annotation files and c) how to publish or integrate annotations within a website without programming. This paper presents the idea and the development of the \textit{semantify.it} platform. It demonstrates that the creation process of semantically annotated data does not have to be hard, shows use cases and pilot users of the created software and presents where the development of this platform or alike projects lead to in the future.
\end{abstract}


\section{Introduction}
\label{sec:Introduction}
The creation of annotations for web content should be neither complicated nor painful, but intuitive and easy for all content creators or web page editors. Not too long ago the challenge was to have a well structured and beautiful looking website. This was solved by the establishment of content management systems (CMS). Now, as the focus on the web shifts away from content- and design based websites towards well structured, high quality content\cite{bizer2008linked, lassila2007embracing} the demand for a CMS like tool to create such structured content grows. 

\begin{figure}
	\centering
	\includegraphics[width=.49\textwidth]{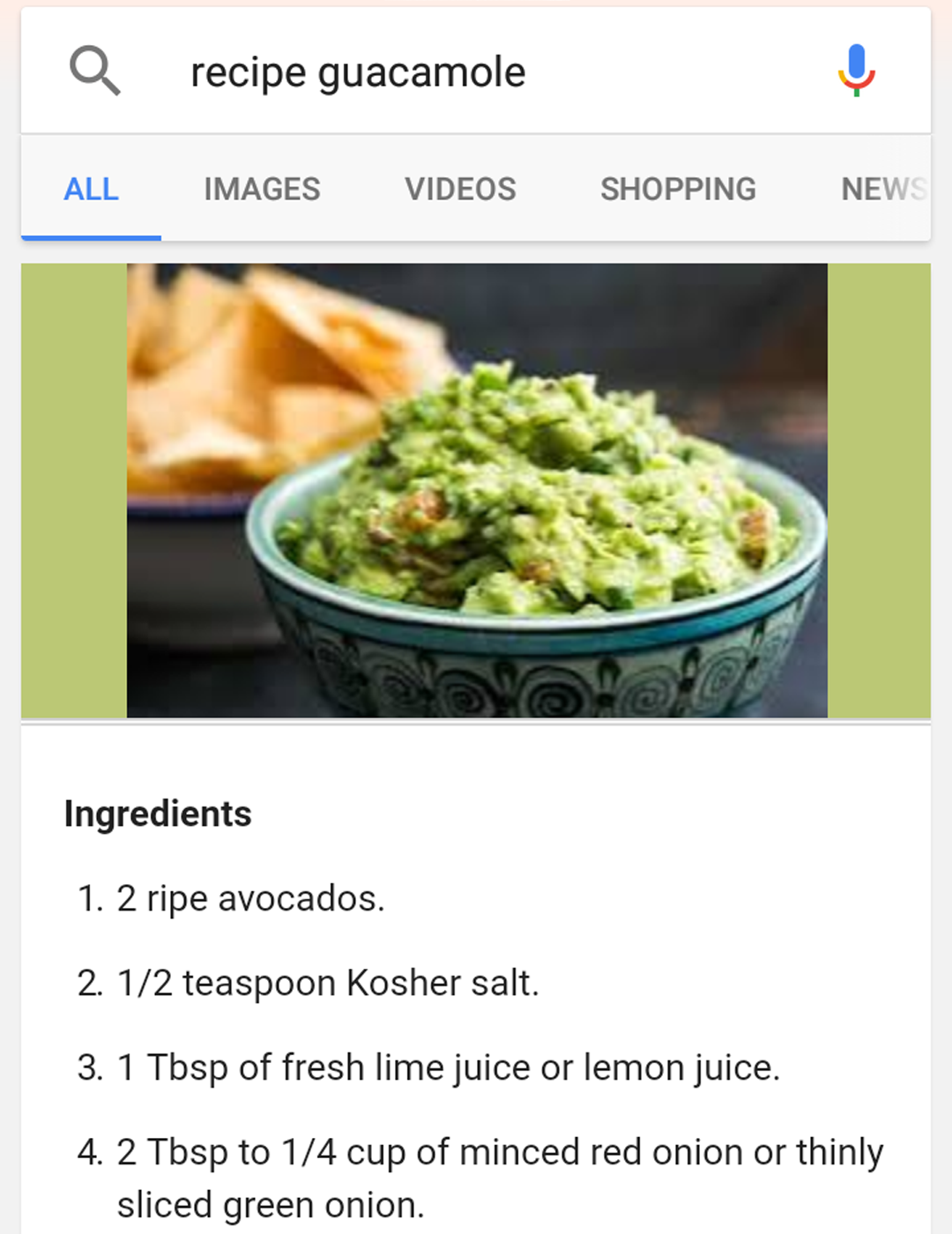}
	\caption{Example of a search for "guacamole recipe" with the result displayed inside the search engine website of Google.}
	\label{fig:google}
\end{figure}

The high demand for annotated data origins in the development of a layer on top of the web as we know it, called the \textit{headless web}\footnote{https://paul.kinlan.me/the-headless-web/}. Within this layer, the number one consumer of content is no longer a human browsing the web, but machines. These machines browse the web with much higher velocity and accuracy and aim to take over search efforts for humans. Intelligent personal assistants (IPA), like Amazon's Echo, Apple's Siri, Google's Allo or Microsoft's Cortana, answer questions, asked by humans, based on high quality structured information from the web. Chatbots too aim towards replacing humans as Q\&A counterparts by retrieving answers from high quality data on the web. The change in the user interface of popular search engines shows that they also try to answer users' demands directly within the search engine website, without the need to lead the user to different, linked, pages. See Figure \ref{fig:google}, for an example, of a search result displayed inside a search engine website.

To structure the content on the web, there is a variety of vocabularys to choose from but the most widely acknowledged one \cite{meusel2015web,guha2016schema} has proven to be schema.org\footnote{http://schema.org}. Schema.org is an initiative launched by the "big 4" search engine providers Bing, Google, Yahoo! and Yandex in 2011. It is a collection of terms to describe \textit{things} on the web in a structured way. It is embedded into the HTML source with either RDFa\cite{adida2008rdfa}, Microdata\footnote{https://www.w3.org/TR/microdata/} or JSON-LD\cite{lanthaler2012using,sporny2014json}. In this work, we will only focus on the latter one.
An analysis conducted by K\"arle et al.\cite{karle2016there} has shown that schema.org is widely distributed, but used mostly in an incomplete or wrong way. But why is the creation of annotations so hard in the first place? The root of the problem can be summarized in three questions: (1) which vocabulary to choose, (2) how to create JSON-LD files and (3) how to publish JSON-LD files.

With \textit{semantify.it}\footnote{https://semantify.it} we provide a web application whose main purpose is to make the creation of annotations easy and intuitive. But it is not only a platform for creation and storage of annotations but also to validate, edit, analyze and publish annotated data.

The web based software is free of charge and anyone can register and start creating annotations based on pre-built forms made by domain experts or uploading annotations. The generator is easy and intuitive to use and the resulting JSON-LD files are stored on the server. From there they can be fetched and integrated into existing websites with the help of content management system plugins or by a unique URL call. Additional to the static data, the platform contains an extension framework, through which applications that map external data sources to schema.org, can push dynamically created annotations to the \textit{semantify.it} platform.

The remainder of this paper is structured as follows: Section \ref{sec:RelatedWork} lists work related to the approach presented in this paper and states the motivation to build the \textit{semantify.it} platform. Section \ref{sec:Methodology} shows the technical approach and Section \ref{sec:Implementation} describes implementation details. Section \ref{sec:Results} presents the results of the work on the platform and Section \ref{sec:Conclusion} concludes the paper and gives an outlook to future work and additional projects.

\section{Related Work \& Motivation}
\label{sec:RelatedWork}
In this Section, we will review the existing annotation tools and frameworks as well as CMS extensions and explain our motivation for developing the \textit{semantify.it} platform to facilate the annotations process.
Annotation of unstructured content on the web has drawn a lot of interest from the semantic web community. Since schema.org emerged in 2011, all parties on the web have gained major motivation for annotating their content, especially for the benefits coming from the support of the major search engines to structured data markup. The recent developments in the intelligent personal assistants (IPA) and chatbots also increased the importance of semantically described structured data on the web. The content on a webpage can be semantically enriched by embedding the annotation of the content to HTML source in formats such as JSON-LD, Microdata and RDFa. However, without proper tool support, the structured content publishing process can be very challenging for the end-user. 

There are many annotation tools and frameworks in the literature with different levels of automation (e.g., automated annotation with natural language processing) \cite{uren2006semantic}. Comprehensive surveys of such tools and frameworks can be found in \cite{Khalili2013review}, \cite{reeve2005survey} and \cite{Gangemi2013}. These annotation and knowledge extraction tools aim to semantically enrich documents and to enable semantic search and reasoning. However, these tools did not find major practical use for annotation of webpages, since they do not create full annotations, but mostly recognize and link entities in text. The technical challenge of embedding annotations into the webpage have been tackled by  extensions/plugins for popular CMS \cite{Khalili2013} \cite{NavarroGalindo2012}. Our approach decouples the generation and publication of the annotations, which allows experts who do not necessarily have access to the administration panel of the CMS to create annotations. Then, our generic CMS extensions can find and inject annotations to webpages. Since the CMS extensions share a common PHP API for communicating with our platform, the CMS specific development effort is kept minimal. Besides the creation and publication, another major challenge of the annotation process remains mostly untouched. Schema.org is a relatively large vocabulary with many types and properties and it is not easy for an end-user to pick relevant types and properties for annotations in a certain domain. Morover, CMS extensions generally support a predefined set of types and properties\footnote{Mostly Article and BlogArticle with mappings from metadata fields of CMS posts to corresponding properties of the aforementioned types.}. An exception could be RDFaCE \cite{Khalili2013}, which allows users to pick desired types from the entire schema.org vocabulary, but the selection is only limited to types, the properties and ranges cannot be restricted. Additionally, with our approach, we enable the creation of annotations based on the frequently changing data, which is not feasible to annotate manually with a CMS extension. The mappings from an APIs data structure or a relational database schema to the schema.org vocabulary should be done. This task requires major development on the CMS side. 

We propose the \textit{semantify.it} platform which facilitates creation, validation and publication of structured data on the web. The annotations can be done manually via an editor that is generated automatically based on a domain specification\footnote{a specific subset of the schema.org vocabulary (see \cite{simsek2017domain} for details)} or automatically through an extension that maps the data structure of external data sources to a domain specification. The data from the mapped data sources then can be pushed to the system via an open RESTful API. Creating annotations against a domain specification (e.g., Google structured data guidelines) helps end-users to ensure that their annotations are compliant with search engines' structured markup guidelines. We are also implementing a rule based validator for semantic validation of the annotations.
The publication of the annotations are done by simple generic extensions that we develop for popular CMS, which merely maps generated annotations to web pages.
Additionally, our open RESTful API allows application developers to reuse the annotations hosted in \textit{semantify.it}, without crawling.

A recent effort from the W3C Web Annotation Working Group, the Web Annotation Data Model\footnote{https://www.w3.org/TR/annotation-model/} and Vocabulary\footnote{https://www.w3.org/TR/annotation-vocab/}, aims to standardize the annotations on the web. The ultimate goal of the standard is to open and decentralize the comments on the web content. It also allows more fine-grained annotations, meaning that it is possible to make comments on a part of the content. This effort does not relate to the purpose of our platform directly, since it is actually a vocabulary for describing the annotations. Nevertheless, the idea of separating annotations from content and publish them on-demand is somewhat parallel to our vision.

To the best of our knowledge, there is no such platform that generates, validates and publishes annotations in a holistic way. By decoupling the annotation creation and publication, we enable content creators who do not have extensive knowledge about schema.org to benefit from semantic annotations, since they can be externally generated by experts and be stored on \textit{semantify.it} platform.

\section{Methodology}
\label{sec:Methodology}
In Section \ref{sec:Introduction}, we introduced the three major challenges when it comes to the annotation of content on the web. The first and probably most important question to answer is, which vocabulary to use. Due to its growing importance and distribution we choose to support the \textit{schema.org} vocabulary\cite{guha2016schema}. But still, there are hundreds of classes and properties to choose from, which makes it very hard for inexperienced users to select the right set of classes and properties to annotate certain web content. To solve that challenge \textit{semantify.it} provides a set of recommendation files for different domains or use cases which define, case specific, which classes and properties are recommended or even required to create proper annotations. Since those recommendation files always target a certain domain they are called \textit{domain specification} or \textit{DS} in short\cite{simsek2017domain}. The second question is how to create proper annotation files in the recommended JSON-LD format\footnote{https://developers.google.com/search/docs/guides/intro-structured-data\#markup-formats-and-placement} (JSON-LD is a JSON-based data format for linked data purposes, which became a W3C recommendation in 2013 \cite{world2014json}). To answer that, \textit{semantify.it} provides the user with specific editors for specific domains or use cases. The editors are based on the selected classes and properties from question one and are generated dynamically every time the user creates an annotation. The third question \textit{semantify.it} addresses is the publication of annotations. Of course, the annotation file can be copy-pasted into a script tag of a website, but most web- or CMS users are not able to fulfill this task. So \textit{semantify.it} stores all created annotations and provides them through an application programming interface (API) and also offers a number of plugins to popular content management systems to automatically retrieve the annotation files from the \textit{semantify.it} server and inject them into the website.

\subsection{Creation}
\label{sec:creation}

\begin{figure}
	\centering
	\includegraphics[width=.47\textwidth]{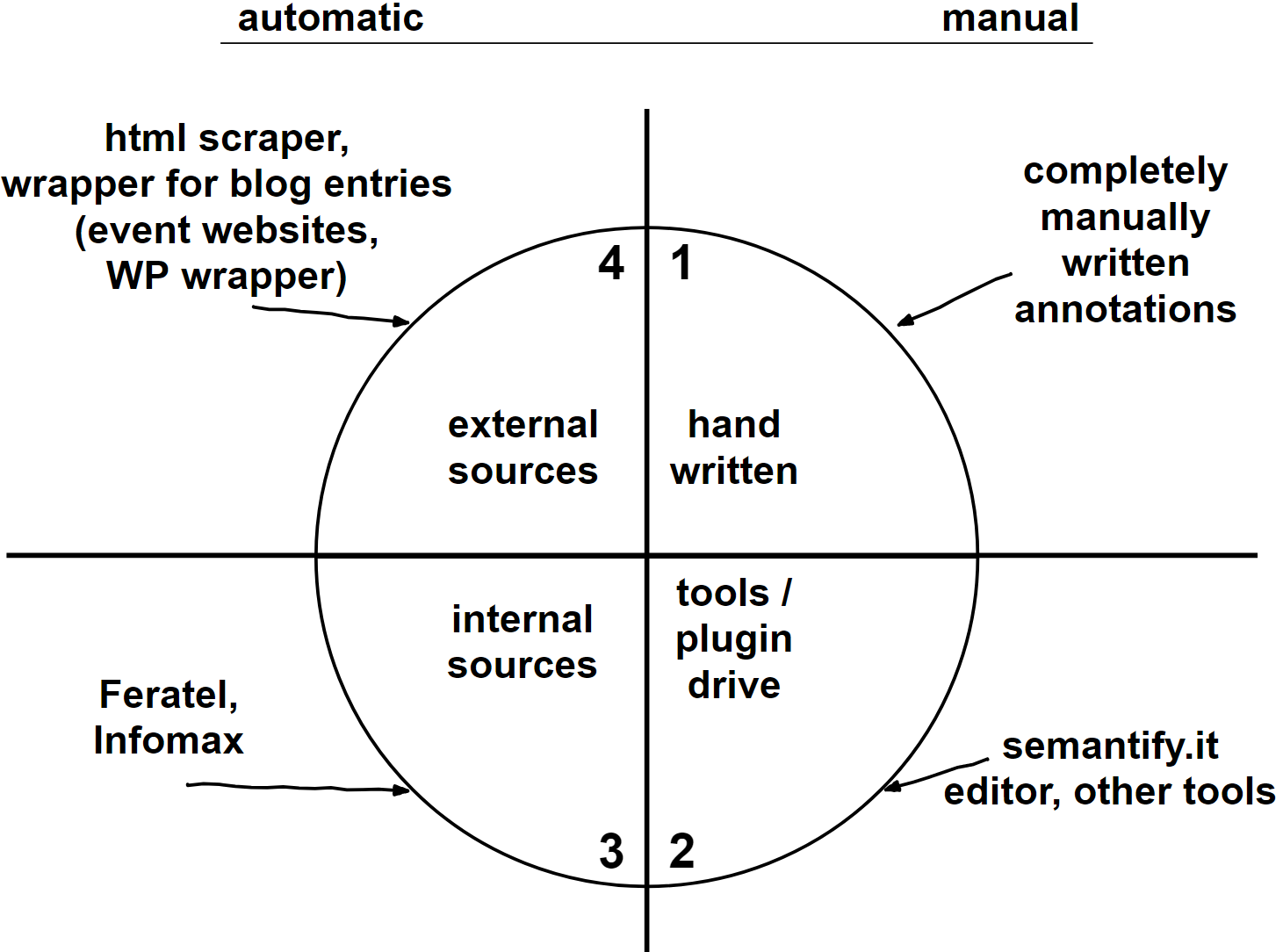}
	\caption{The four different types of content creation. Every quarter stands for a different type how an annotation can be created.}
	\label{fig:circle}
\end{figure}

As depicted in Figure \ref{fig:circle}, we distinguish between manual and automatic annotation creation. The two options have two more distinctions. Manual annotations can either be completely handwritten, with a text editor (first quarter), or tool driven, like with the \textit{semantify.it} editor (second quarter). Automatic annotations can be divided, from a service point of view, into internal sources and external sources. We talk about internal sources if the structure of the data is known or agreed on and the structure is maybe even protected by a service level agreement (third quarter). External sources are such, where the structure is unknown and has to be determined based on an HTML source. The structure can change any time and no agreement between the data provider and the annotation creator exists (fourth quarter).

\subsubsection{Manual annotation creation}
The manual annotation creation process is, as mentioned above, driven by editors based on domain specifications. The concept of defining subsets of schema.org annotations for domain specific usage was first presented by \c{S}im\c{s}ek et al. in \cite{simsek2017domain} and was adapted for the usage in \textit{semantify.it}. To build the DS files, the platform features an editor, which lets the domain expert select classes, properties and sub classes just by clicking. No source code has to be written at all. The DS files are then saved and accessible to all users over the annotation editor interface. When starting a new manual annotation, the user selects a DS on which the annotation will be based on and gets presented with the corresponding editor. The editor for annotation creation looks like an ordinary HTML form and hence gives the user a good and familiar usability. If all required fields are filled the user can proceed by clicking the "generate" button and gets presented with the annotation source code in JSON-LD format. This source code can then either be copy-pasted or stored on the \textit{semantify.it} platform for further usage.

Manual creation of annotations with an editor is only one way to use \textit{semantify.it}. Of course the platform does not cover domain specification for all use cases. So someone might create annotations in a different way but still wants to utilize \textit{semantify.it} for storing and distributing those annotations. For this case there is an upload functionality where one or multiple annotations can be posted to the platform where they then are treated exactly as the annotations created with an editor.

To introduce further ways of deploying annotations to the \textit{semantify.it} platform, we first have to define a distinction between three different types of content, which are \textit{static content}, \textit{dynamic content} and \textit{active content}. Static content hardly changes after having been produced once. For example on a hotel website it is mainly the hotel's core data: name, address, phone number, email and alike. Dynamic data changes frequently or even constantly. To stick with the hotel example: room availabilities, prices or specialties on the daily menu count as dynamic data. Active data is information about interfaces to interaction software on a website, like, for example, an internet booking engine's API. The manual creation of annotation with the \textit{semantify.it} editor targets mostly static data. Due to its nature it makes hardly any sense to annotate dynamic data manually.

\subsubsection{Automatic annotation creation}
For the use with dynamic data, \textit{semantify.it} offers, similar to the upload functionality, the functionality to send annotation files to an API to be stored on the platform. So annotations created automatically, for example by a wrapper software, can be stored on and distributed by \textit{semantify.it} as well. Some of those wrappers are even integrated into the platform and provided to the user as \textit{extensions} (see \ref{sec:extensions}). To make use of such a wrapper extension, the user has to activate the extension, provide it with its credential to the given data source and set the frequency for the wrapping process. Then the data of the source, a WordPress blog or a destination management software, is crawled, mapped and stored on \textit{semantify.it} recurrently.

\subsection{Validation}
Another part of annotation creation is the annotation validation. \textit{Semantify.it} offers a validation feature based on the ideas mentioned in \cite{simsek2017domain} to give the user feedback if the information he was entering makes sense. The validation process performs a semantic validation where, based on validation rules, a check is performed if the data entered makes sense according to the rules defined by a domain expert. So for every DS mentioned above, there is the option to create a corresponding domain rule file, or DR for short, to perform semantic validation. Currently the prototypical validation feature is being improved on the platform. The developments on the rule editor are ongoing.

\subsection{Storage \& Publication}
\label{sec:storagepublication}
As mentioned above, most content creators are not able to copy-paste annotation files into their content management systems. So \textit{semantify.it} provides an infrastructure for storage, maintenance, analysis and publication of annotations.
Every file created, uploaded or stored through the public API, is assigned to a concept called \textit{website}. A \textit{website} is associated with a \textit{user} who can create several of those \textit{websites}. Every \textit{website} has an API key, which is used to fetch annotations from or store annotations to said \textit{website}. Mostly the \textit{website} concept of \textit{semantify.it}, as a collection of annotations, correlates with a real website where the annotations belong to that is the reason for the naming. An annotation is uniquely identified by a nine alphanumeric character long URL safe hash code. To retrieve the annotation from the \textit{semantify.it} server the user just has to enter the shortener URL\footnote{https://smtfy.it/} and append the hash code. The response is a plain JSON file containing the corresponding annotation in JSON-LD format.
On the dashboard the platform shows all annotations grouped by \textit{website}. Every annotation has the possibility to be previewed or edited. Editing works by loading the corresponding editor, populating the form with the content from the annotation and overwriting the old annotation when the user is done editing. 
\textit{Semantify.it} provides an analytics feature for (so far) basic statistics about the number of classes and properties annotated and the overall number of facts stored for each \textit{website}. This functionality will be extended in the future (see Section \ref{sec:Conclusion}).
There are several ways to publish annotations stored on \textit{semantify.it}. For static data it might make sense to fetch every annotation separately for a webpage by the hash identifier. For dynamic data there are two possibilities: (1) if an annotation gets stored on \textit{semantify.it}, the software checks for a valid schema.org/url property. If that property exists it gets URL-encoded and stored as a retrieval key for that annotation file. The annotation can then be fetched by calling the shortener URL followed by "url/" and the URL-encoded content of the schema.org/url parameter. This method makes sense when a web master decides to automatically annotate a huge number of blog entries and store them on \textit{semantify.it}. The annotations can then be retrieved with a CMS plugin (see \ref{sec:plugin}) where each annotation file is identified by its encoded URL. (2) with a custom identifier, called CID. The API call to send annotations to \textit{semantify.it} offers the possibility to add an optional CID parameter for each annotation. Annotations stored with a CID can be fetched from the server by calling the shortener URL followed by "cid/" and the value of the custom identifier. This makes sense for systems where web content is stored in a database and then injected into a web page based on a CID. Those web pages can be annotated automatically and the annotations can be injected, just as the web content itself, by the corresponding CID. As part of the publication functionality, \textit{semantify.it} provides a number of plugins for popular content management systems (so far for WordPress\footnote{https://wordpress.com/} and Typo3\footnote{https://typo3.org/}, but Joomla\footnote{https://www.joomla.org/}, Drupal\footnote{https://www.drupal.org/} and more are in the pipeline). Those plugins can be downloaded from the CMS provider's plugin repository. Therefore, the \textit{website's} API key has to be stored in the plugin's settings and the configuration has to be set to either load the annotations manually per web page or automatically by web page URL. Then, on every web page call, the annotations get fetched from \textit{semantify.it} and injected into the web pages created by the CMS.  

\subsection{Extensions}
\label{sec:extensions}
Besides the possibility to create annotations manually and to use the service as a storage, maintenance, analytics and publication platform, \textit{semantify.it} also offers an extension functionality, which targets automatic annotation creation. An extension is actually a piece of standalone software, which is integrated into the \textit{semantify.it} platform. A user can optionally activate extensions and configure them individually. Extensions are developed by the \textit{semantify.it} team or can be suggested by external developers through Bitbucket\footnote{https://bitbucket.org/semantifyit} or Github\footnote{https://github.com/semantifyit}. Some examples for extensions are listed below.

\subsubsection{Data mapping} a lot of websites obtain their content from external sources, which have public APIs. For example a destination management organization's (DMO) website contains data about room offers or hiking paths and the data is provided by different vendors through their APIs. If the DMO wants the content to be annotated it either has to convince the data provider to annotate all the data (which is probably hard) or use the corresponding \textit{semantify.it} extension. The extension requires the API access data of the user and then starts to crawl the data, map it to schema.org and store the annotations on \textit{semantify.it}. A simple plugin can then pull the annotation from \textit{semantify.it} and inject the right annotation to the corresponding web page. An example for the use of data mappings for massive annotations of destination management organizations' websites can be found in \cite{akbar2017massive}.

\subsubsection{WordPress article annotation} another example for an extension is the annotation of blog articles in WordPress. Currently there are no plugins which annotate pages or blog entries in WordPress directly. So, if an author decides to create annotations for all his old articles this can become very painful. So \textit{semantify.it} provides an extension, which crawls all blog articles of a given website, maps the relevant content to schema.org and stores the annotation file on \textit{semantify.it}. A plugin, like the one explained above, can then fetch the annotation and inject it into the corresponding WordPress web page. The same could work for other blog systems too.

\section{Implementation}
\label{sec:Implementation}
\textit{Semantify.it} was designed and implemented to be delivered as a software as a service or SaaS \cite{buxmann2008software}. 
To support version control during the development we make use of the free and open source version control software Git\footnote{https://git-scm.com/}, hosted on the platform Bitbucket. Through short commit cycles, sophisticated branching and meaningful commit messaging the code is kept as manageable as possible, relatively easy to maintain and easy to roll back in case grave errors should be detected only after a release.

\begin{figure}
	\centering
	\includegraphics[scale=0.13]{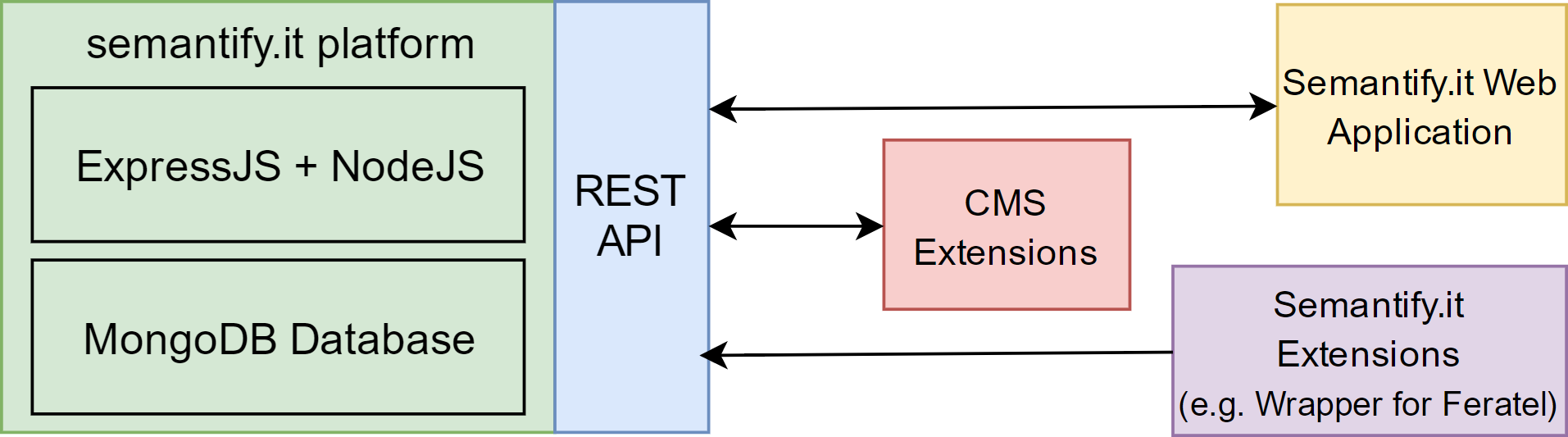}
	\centering\caption{Modules of the \textit{semantify.it} platform.}
	\label{fig:architecture}
\end{figure}

The reference architecture of the \textit{semantify.it} is depicted in Figure \ref{fig:architecture}. In the following subsections, we will explain the main modules in our architecture and the communication between external applications and the \textit{semantify.it} platform.

\subsection{Platform Core} 
The platform core has been implemented with NodeJS and the ExpressJS web application framework, which allowed us to create a lightweight web based platform with a RESTful API. We adopt the document-oriented database system MongoDB for persistance. A document-oriented database is a natural decision when working with JSON-LD files, since they can be stored directly as documents. This way, we can serve the annotations as they should appear in the HTML source of a web page. Another possibility such as using a triple store would be more suitable for developing semantic web applications based on the stored annotations, however this would make querying the annotations to obtain JSON-LD documents to be embedded into a web page more challenging. Even though it would be possible to query a single blank node and all the other nodes that are connected to this blank node with the help of property paths, this would still be tricky since an annotation is typically a RDF graph, which consists of many blank nodes connected to each other. Therefore, referring to a specific node without preprocessing the annotations would not be possible. This gets even more complicated when an annotation contains two disconnected graphs. In this case storing each annotation in their own named graph may be a solution. Nevertheless, the main purpose of the \textit{semantify.it} platform is to serve the annotations to web pages, therefore we handle the annotations as JSON-LD documents, rather than graphs. The hosted annotations can be then retrieved via the REST API and stored in a triple store in a desired way. 

The platform manages the annotations in relation with websites, organization and users. We define the concept of website in our data model, which can host multiple annotations. Every website has a unique API key. An organization can have multiple users and they can manage multiple websites that belong to their organizations. External applications can perform CRUD operations on the core platform via the RESTful API. The API key of a website is used by external applications for creating and retrieving annotations. More sensitive operations, such as updating and deleting annotations require additional security measures. In this case, the authentication of users is handled by JSON Web Tokens (JWT).

\subsection{Web Application}
\label{sec:WebApplication}
The web application is developed as an interface to the core platform. In the frontend, the application uses HTML5, CSS, Javascript and Material Design\footnote{https://material.io/} elements. It communicates with the RESTful API of the platform core with JQuery.  The application currently supports the fundamental functionality such as user registration, website management, annotation creation based on domain specific editors, domain specification and bulk upload of annotations. Additionally, users can see certain statistics about their websites (e.g., number of annotations, number of statements, number of annotation requests). The web application has access to all the routes defined in the RESTful API.

\subsection{CMS Plugins}
\label{sec:plugin}
We develop two plugins for the popular CMSs Wordpress and TYPO3. According to Web Technology Surveys\footnote{https://w3techs.com/technologies/overview/content\_management/all}, Wordpress is the most widespread CMS worldwide, and TYPO3 is very common in German speaking countries. Therefore, plugins for these CMS in the initial phase covers many use cases. Both extensions use a common PHP API to communicate with the RESTful API of \textit{semantify.it} platform. The front-end development of the plugins is tailored for each CMS since they vary in plugin architecture. The only configuration the CMS plugins need is the API key of a website on \textit{semantify.it}. The plugins have two main functionalities; (a) they allow the content creator to tie a specific page/post with a specific annotation hosted in the \textit{semantify.it} platform and (b) the plugin can use the URL of a page and retrieve an annotation that has the same URL as the value of the \textit{url} property. This feature is very useful in most cases, however, a user can always opt out from using it and select annotations manually.

\subsection{Extensions}
One of the most important challenges of semantic annotations on the web is the maintenance of them. In many cases, the important data usually changes frequently, therefore keeping the annotations up to date is an important task. For instance in the tourism domain, accommodation offers can change on a daily basis. In the \textit{semantify.it} platform, we offer an extension mechanism where the data from external data sources (e.g., Feratel) are mapped to domain specifications and annotations are generated automatically with a specified frequency. Annotation of frequently changing data through wrappers has been described in \cite{akbar2017massive}. We also create wrappers as extensions to the \textit{semantify.it} platform, which can be activated by users when needed. The mapping is currently done within the wrapper's business logic, but we plan to adopt an RDF Mapping Language (RML) \cite{dimou2014rml} based approach  in order to increase re-usability (see Section \ref{sec:Conclusion}).
The automatically generated annotations are stored in the database with a unique custom ID (CID). This ID is generated based on the external data source's entity identification scheme. For instance, Feratel uses UUID\footnote{https://en.wikipedia.org/wiki/Universally\_unique\_identifier} for identifying entities such as accommodations. Based on this, we create CIDs in "FeratelID-languageCode" format, since we want to identify annotations in different languages separately. The annotations then can be matched on the client's side with the corresponding webpage where the content about an entity resides in a certain language.

The main challenge of the extension development lies in the mapping of custom data formats and structures to schema.org vocabulary. In some cases the entity types in the external source's data model may be too specific for schema.org vocabulary. We overcome this challenge by using a suitable more generic type from schema.org. Another challenge is that some information may be given in an unstructured way in the external data source, which makes it tricky to map and extract programmatically. In such cases, we try to find patterns in the content and write suitable extractors. If this is not possible, we simply ignore that content.  

\section{Results \& Use Cases}
\label{sec:Results}
This Section will show the use cases of our implementation and present statistics obtained from the initial usage.

The platform started with one ski school as a pilot. Meanwhile the three destination marketing organizations (DMO) of Mayrhofen, Seefeld and F\"ugen are testing \textit{semantify.it} as pilots and the umbrella organization of Tirol's tourism organization, Tirolwerbung\footnote{http://tirolwerbung.at}, is about to use \textit{semantify.it} with a wrapper extension. Besides that, several private websites are working with \textit{semantify.it} and provide feedback. A more detailed description of those use cases will be presented in \ref{sec:usecase}. Also, the WordPress\footnote{https://wordpress.org/plugins/semantify-it/} and the Typo3\footnote{https://typo3.org/extensions/repository/view/semantify\_it} plugin are used already by pilots and deliver thousands of annotations every day, Section \ref{sec:results} gives more details about that. 

\subsection{Results}
\label{sec:results}
At the time of the evaluation, \textit{semantify.it} was hosting 31 users in 27 organizations maintaining 42 websites. There were 37.597 annotation files stored, containing more than three million annotation statements (triples), which where requested over the API more than 82.000 times in the time span between April 5\ts{th}, 2017 and June 14\ts{th}, 2017. For the better understanding of the annotation file to annotation request ratio, it is important to mention that not all pilots test the whole work flow of \textit{semantify.it}. Some are testing the bulk upload feature through data wrapper extensions, which leads to a huge number of annotation files, but the CMS extension not yet, which explains the relatively low annotation request number. Others created their annotation files manually with the \textit{semantify.it} editor or a text editor but use the CMS extension, which leads to only small number of annotation files but a big number of annotation requests. Every page call on the CMS extension user's website triggers one annotation request on the \textit{semantify.it} platform. So only the pilots having installed the CMS extension contribute to that number. As soon as all pilots use any form of CMS extension, the number will increase drastically. Currently the Typo3 plugin counts 127 downloads (which are not unique per website) and the Wordpress plugin counts less then 10 active installs.

\begin{figure*}
	\centering
	\includegraphics[width=.99\textwidth]{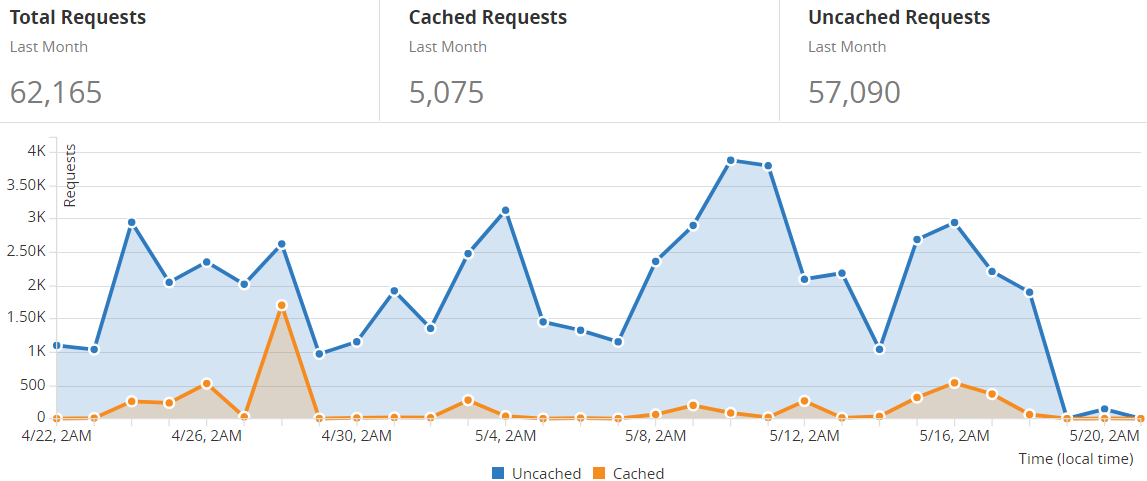}
	\caption{Access statistics of \textit{semantify.it} for the one month time span from April 22\ts{nd} to May 20\ts{th}.}
	\label{fig:cloudflare}
\end{figure*}

To provide SSL\footnote{Secure Sockets Layer} capabilities to the users, \textit{semantify.it} traffic is channeled through Cloudflare\footnote{https://www.cloudflare.com/}, a content delivery network. A picture of Cloudflare's analytics service shows the accesses to \textit{semantify.it} (over UI as well as over the API) in the time from April 22\ts{nd} to May 20\ts{th} (see Figure \ref{fig:cloudflare}).

To find out if loading annotations for websites from the \textit{semantify.it} platform is acceptable in terms of response time we performed a series of response time measurements over a testing website\footnote{https://tools.pingdom.com}. The average response time was at around 150ms, which is an acceptable loading time for external scripts.

\subsection{Use Cases}
\label{sec:usecase}
To test the functionality and the operational readiness we applied several different use cases to \textit{semantify.it} and tried to find pilots for all four annotation creation scenarios described in Figure \ref{fig:circle}. We created and uploaded annotations manually and distributed them via the CMS extensions and we used annotations, which were created automatically and were uploaded to \textit{semantify.it} vie the API. Those scenarios will be described below.

\subsubsection{Manually created annotations}
The first pilot of \textit{semantify.it} was a ski school from Switzerland. Their website consists of 64 sub pages of static, rarely changing, content. For the purpose of being a \textit{semantify.it} pilot, all annotations were created manually and uploaded through the platforms upload-feature. The total count of annotation statements in all the 64 annotation files is 5312, which means that there are 5312 facts or triples stored on \textit{semantify.it}. The website uses a Typo3 CMS and has the \textit{semantify.it} plugin installed. The administrators went through all the 64 sub pages and selected the corresponding annotations manually. This use case matches scenario one in Figure \ref{fig:circle}.

A use case for scenario two from Figure \ref{fig:circle} was a hotel pilot. The annotations for the hotel, the included restaurant and some events were made with the \textit{semantify.it} editor and integrated into the hotel's website with the Wordpress plugin.

\subsubsection{Automatic annotation creation}
A use case for automatic annotation creation (Figure \ref{fig:circle}, third quarter) was the mapping of Feratel's\footnote{http://www.feratel.at/loesungen/feratel-destination/datenmanagement/web-client/} tourism destination data into schema.org (as described in \cite{akbar2017massive}). The thousands of annotation files of the DMOs of Mayrhofen, Seefeld and F\"ugen, first stored in a file system, made a perfect use case for \textit{semantify.it}. So we extended the existing wrapper and now every night all the data for the corresponding website from the Feratel system, annotated with schema.org, is uploaded to \textit{semantify.it}. For the three DMOs mentioned above there are currently around 22,000 annotation files containing 3.9 million annotation statements. The annotation files, identified by a UUID stored as CID, are replaced if they already exist, otherwise newly created. The CMS plugin, which is not made by us but by the DMOs' web agencies, is not ready yet. But we could find out that \textit{semantify.it} can easily cope with thousands of annotation files and millions of annotation statements and the performance of the upload API scales.

A similar use case is the example of Tirolwerbung. Their web agency maintains a self made CMS with an API to the database. We built wrappers for various different domains (hiking routs, ski resorts, accommodations and others) and now daily crawl the API to then store the resulting annotation files (around 6,000) on \textit{semantify.it}. As in the previous example the annotation retrieval software for the CMS is not yet finished. The annotation files are identified by a CID with which they are also going to be fetched by the CMS plugin.

Another use case is the annotation of a corporate blog with around 220 entries. As an example for scenario four in Figure \ref{fig:circle} we wrote a script, which scraped the content, mapped it to schema.org and stored it on \textit{semantify.it}, which led to 14,191 annotations statements in 223 annotations files. The blog is built on Wordpress and through the use of the plugin the annotations are injected into the blog's HTML. In this use case the automatic annotation retrieval by URL property (described in \ref{sec:storagepublication}) comes into play. Thus, every annotation is integrated into the blog automatically and the administrator does not have to spend time assigning annotation files to web sites.

\section{Conclusion \& Future Work}
\label{sec:Conclusion}
This Section concludes the work on the\textit{semantify.it} platform, wraps up the outcome and gives an outlook into the future of developments on the software.

\subsection{Conclusion}
This paper describes the \textit{semantify.it} platform, a multi purpose software-as-a-service to create, store, validate, publish and analyze semantic data. The easy-to-use interface and the comprehensive API make it easy to generate and store semantic annotations. The extension system, which generates annotated data out of different data sources makes annotations even more accessible for different users and purposes. Plugins to popular content management systems make the usage and publication of this structured data simple for non experienced users. Even though the individual parts of \textit{semantify.it} might not be complete novelties, the idea of a holistic platform for creation, publication and distribution of semantic annotations is novel. As a proof-of-concept, use cases from the tourism field in Austria, Germany and Switzerland show that the \textit{semantify.it} platform is capable of handling real life traffic reliably. 

\subsection{Future Work}
There is still a lot to be done to make the creation and publication of semantic annotations easy and intuitive. Our efforts in enhancing \textit{semantify.it} go in several directions, which will be shortly described below. To ensure the usability of the \textit{semantify.it} user interface we are about to conduct a usability study, which might lead to further improvements and a qualitative comparison of \textit{semantify.it} with other annotation tools.

\subsubsection{Incoming data processing} In the future, we will put a lot of emphasis on processing of data, which is already structured, by a database or an API, but not annotated. For that we are planning on enhancing our extension system (see \ref{sec:extensions}) towards being more flexible and generic. To reach that we plan on integrating an RML processor\cite{dimou2014rml} and providing templates to easily describe data sources in RML. This will improve the work flow of integrating new structured data sources a lot and help to provide more incoming data sources for \textit{semantify.it} users.

\subsubsection{Advanced validation}
To improve data quality of annotations we will provide more advanced validation measures to the annotation editor and validation mechanisms to the upload and API interfaces. Based on the ideas presented in \c{S}im\c{s}ek et al.\cite{simsek2017domain} we will provide a rule designer interface and a set of ready made rules to support users of \textit{semantify.it} in generating and storing only semantically valid, high quality data.

\subsubsection{Advanced analysis and reasoning}
For more ambitious users of \textit{semantify.it} we will provide an improved analytics feature. This functionality will let users see statistics about the number of annotations they made, how much classes and properties they are using and detailed statistics about how often and by whom their annotations are fetched. This will provide users with a better inside into who consumes their data and will hence lead to better annotations.
And since the \textit{semantify.it} platform stores a huge amount of public accessible, structured, high quality data we think of also setting up a over-all statistics website with anonymous insides into what data is available and the performance of \textit{semantify.it}'s user's websites. The data can maybe also be used, anonymized of course, to make predictions about certain fields in which a lot of websites use \textit{semantify.it} for annotations.
As the part of our future work, we will create a knowledge graph for tourism in Tyrol region in Austria by exporting relevant annotations from the \textit{semantify.it} via the REST API and loading them into a triple store after preprocessing (e.g., identification, reconciliation). This will help us to apply reasoning and reveal the implicit knowledge. Additionally, with the help of historical data, we will be able to apply daata analytics such as showing the price trend thoughout a year in a region. 

\subsubsection{On thy fly adaptability to schema.org versions}
The consortium behind schema.org tries to drive development by releasing new versions of schema.org in relatively short cycles. The updates mostly feature significant changes to the core vocabulary. In version 3.1, for example, 12 classes and 10 properties for accommodation businesses were introduced (as described in \cite{karle2017extending}). To keep \textit{semantify.it} always up do date we are going to implement an on-the-fly adaptability feature where, whenever a new schema.org version is released, the \textit{semantify.it} editor uses the latest vocabulary from the Github repository of schema.org.

\section*{Acknowledgment}
The authors would like to thank the Online Communications working group (OC)\footnote{http://oc.sti2.at/} for their active discussions and input during the OC meetings, our colleague Oleksandra Panasiuk for the creation of the domain specifications for showcase domains on \textit{semantify.it}, our colleague Zaenal Akbar for the adaption of the Feratel wrapper, our colleague Boran Taylan Balci for the programming of the corporate blog scraper and the \textit{semantify.it} development team (in the order of joining the team) Omar Holzknecht, Roland Gritzer, Richard Dvorsky and Dennis Sommer, for their hard work and their commitment to the mission of making the semantic web real and usable for everyone. A special thank you goes to all the testers and pilots who were giving \textit{semantify.it} a chance, regardless the missing and often buggy features along the way. There is no good product without good testers!


\bibliographystyle{IEEEtran}
\bibliography{sections/bib}

\end{document}